\documentclass[twocolumn,showpacs]{revtex4}
\usepackage{amsmath}
\usepackage{graphicx}% Include figure files

\begin{document}

\title{Spin waves in quasi-equilibrium spin systems.}

\author{Kevin S. Bedell}
\affiliation{Department of Physics, Boston College, Chestnut Hill,
 MA, 02467}

\author{Hari P. Dahal}
\affiliation{Department of Physics, Boston College, Chestnut Hill,
    MA, 02467}

\date{\today}

\begin{abstract}
Using the Landau Fermi liquid theory we have discovered a new
regime for the propagation of spin waves in a quasi-equilibrium
spin systems. We have determined the dispersion relation for the
transverse spin waves and found that one of the modes is gapless.
The gapless mode corresponds to the precessional mode of the
magnetization in a paramagnetic system in the absence of an
external magnetic field. One of the other modes is gapped which is
associated with the precession of the spin current around the
internal field. The gapless mode has a quadratic dispersion
leading to some interesting thermodynamic properties including a
$T^{3/2}$ contribution to the specific heat. We also show that
these modes make significant contributions to the dynamic
structure function.
\end{abstract}

%%% ----------------------------------------------------------------------
\maketitle
%%% ----------------------------------------------------------------------
There has been much renewed interest in studying non-equilibrium
(NEQ) spin polarized paramagnetic Fermi liquids and gases. The NEQ
spin system can be defined as a polarized spin system which can be
obtained by maintaining an unequal population of different spin
species to create an off-balanced Fermi level in the absence of an
external field. This is both an old
subject\cite{bouchiat1960,colegrove1963}, and a rather new
one\cite{leggett2001,ketterle2002,martin2005,zutic2004,cho2005}.
The NEQ spin systems have been created in a wide variety of
materials using many novel techniques. An example of the most
recent one would be the creation of steady state spin polarization
in a system of cold and ultra-cold Fermi gases of alkali metals.
In these system one uses optical trapping and atomic manipulation
to get unequal populations of the fermions in two different
states\cite{leggett2001,ketterle2002,martin2005}. The NEQ spin
systems have also been created in many materials which comprise
the study of one of the most promising areas of physics, namely
spintronics\cite{zutic2004,fernandez2004,malshukov2005,yaroslav2002}.
In this area of study a couple of methods are in practice to
achieve the NEQ state of these spin systems.

Widely used methods for achieving NEQ states are optical pumping
and polarized spin injection. In the optical pumping
method\cite{zutic2004}, electron orbital momentum is directly
oriented by a strongly circularly polarized laser light and
through the spin-orbit interaction the electron spin is polarized.
In the polarized spin injection\cite{zutic2004} method one
connects a magnetic electrode to a normal metal sample and drives
a polarized current using external biasing which creates unequal
population of spins in the normal metal. In such a ferromagnet -
metal bilayer structure, it has been suggested that if the
direction of the magnetization in the ferromagnet is made to
precess (by applying an external magnetic field on the
ferromagnet) we can inject polarized spin into the normal
metal\cite{yaroslav2002}. It has also been suggested in the
literature that we can use the interaction of the electron spin
with the strain-induced potential to create the NEQ system in a
nano-mechanical system\cite{malshukov2005}. The NEQ system of
polarized liquid helium is one example of an old but still
interesting subject\cite{candela1994}. The rapid melting method
suggested by Castaing and
Nozieres\cite{castaing1979,chapellier1979}, and the optical
pumping\cite{bouchiat1960,colegrove1963,candela1994} were widely
used to create the NEQ systems of liquid helium.

In this Letter we are interested in studying the thermodynamics
and the spin dynamics of the NEQ systems in the limit of small
spin polarization. We know that for a paramagnetic spin system of
free fermions in the presence of an external magnetic field, the
magnetization, $m$, precesses around the applied magnetic field at
the Larmor frequency. For a small transverse perturbation $\delta
\overrightarrow{B} = B_x \hat{x} + B_y \hat{y}$, and defining
$\delta B^\pm = B_x \pm  i B_y$ and $m^\pm = m^x \pm i m^y$, the
evolution of $m$ for non interacting Fermi system can be written
as\cite{bedell1993},
\begin{equation}
\frac{\partial m^\pm}{\partial t}= 2i m^\pm \gamma  B
\end{equation}
where $m = m_0 e^{-i\omega_0^\pm t}$ and $\omega_0^\pm = \pm
2\gamma B$ is the Larmor frequency and $B$ is the external
magnetic field. The fact that the magnetization precesses around
the external field with the Larmor frequency also holds true for a
system of interacting fermions, with spin conserving interactions.
Now, for the NEQ systems, since the external field is zero, the
precessional mode should be gapless. To understand the details of
the dispersion of the gapless mode we need to study the
thermodynamics and the dynamics of a Fermi liquid away from
equilibrium.

The general theory of Fermi liquids and gases far from equilibrium
is beyond our current understanding. We will develop here the
theory for weakly spin polarized paramagnetic Fermi liquids with
$B=0$. In this limit we can argue that the Fermi liquid will be in
a quasi-equilibrium(QEQ) state. It is well known that the
polarization dependent chemical potential has the form $\mu(m) =
\mu(0) + O(m^{2})$\cite{carlos1989}, hence for weak polarization,
i.e. for $\frac{m}{n}\ll 1$ where $n$ is the particle density,
$\mu(m) \approx \mu(0)$. This suggests that for small $m$ the
system will be in a QEQ phase and we can study the dynamics of the
spin system around this phase. We know that the variation of the
energy per-unit volume due to a variation in the distribution
function from the ground state is written
as\cite{baym1991,pines1966,carlos1989}:
\begin{equation}
\varepsilon =   \sum_{p \sigma}\varepsilon_{p \sigma}^0 \delta
n_{p \sigma} + \sum_{p \sigma, p'\sigma'}f_{pp'}^{ \sigma
 \sigma'}\delta n_{p \sigma}\delta n_{p' \sigma'}+....
\end{equation}

So the quasi-particle energy can be written as:
\begin{equation}
\label{energy} \varepsilon_{p \sigma} =   \varepsilon_{p \sigma}^0
+ \sum_{p' \sigma'}f_{pp'}^{ \sigma
 \sigma'} \delta n_{p'
\sigma'}+....
\end{equation}
where for a spin conserving interaction with $\frac{m}{n}\ll 1$,
\begin{equation}
\label{lip} f_{pp'}^{\sigma\sigma'} \simeq  f_{pp'}^s +
f_{pp'}^{a}\sigma \cdot \sigma'
\end{equation}

Using Eq.(\ref{energy}) and expanding $f_{pp'}^{s,a}$ in Legendre
polynomials we can derive that the chemical potential of the up
and the down spins will have the forms $\mu =
\varepsilon_{F\uparrow}^0 + \frac{F_{0}^{a}m}{N(0)} + O(m^2)$ and
$\mu = \varepsilon_{F\downarrow}^0 - \frac{F_{0}^{a}m}{N(0)} +
O(m^2)$, where $N(0)=\frac{k_{F}m^{*}}{\pi^{2}}$ is the $m=0$
density of states of the quasi-particles at the Fermi level. So in
leading order in $m$, $\varepsilon_{F\uparrow}^{0} -
\varepsilon_{F\downarrow}^{0} = \frac{-2F_{0}^{a}m}{N(0)}$. For
the EQ system, $\mu = \varepsilon_{F\uparrow} - B +
\frac{F_{0}^{a}m}{N(0)} + O(m^2)$ and $\mu =
\varepsilon_{F\downarrow} + B - \frac{F_{0}^{a}m}{N(0)} + O(m^2)$.
Again in leading order in $m$, $\varepsilon_{F\uparrow} -
\varepsilon_{F\downarrow} = 2B - \frac{2F_{0}^{a}m}{N(0)} =
\frac{2m}{N(0)}$. Contrary to the EQ system, we see that the
change in the Fermi energy of the up spin and the down spin
fermions of the QEQ system depends explicitly on the Fermi liquid
parameter $F_0^a$. For the QEQ and EQ systems with a majority of
up spins, we have shown, in Fig.1a, the change in the Fermi energy
of the up spin ($\varepsilon_{F\uparrow}^0$) and the down spin
($\varepsilon_{F\downarrow}^0$) fermions, and in Fig.1b, the
energy spectrum of these spin species. Here $B_{eff}$  is the
effective field. For the QEQ system $B_{eff}=\frac{-mF_0^a}{N(0)}$
and for the EQ system $B_{eff}= \frac{m}{N(0)} =
\frac{B}{1+F_0^a}$. For a QEQ system, Fig.1a corresponds to a
steady state with constant number density which can be obtained,
for example, either by taking out some down spins and putting in
equal number of up spins or by flipping the down spin over. To
maintain such steady states we need to continuously supply energy
to the system (e.g., continuously shining light in the system in
the optical pumping method) against the loss in the polarization
due to spin flip scattering mechanism. In what follows we want to
fix the number density of the Fermions to avoid any complication
that would be introduced due to the coupling between the charge
density fluctuations and the spin density fluctuations.

To study the collective excitations of the QEQ system we
investigate the  oscillations of the transverse component of the
magnetization ($\delta \overrightarrow{m}_{p}$). The linearized
kinetic equations for the evolution of $\delta
\overrightarrow{m}_{p}$ can be written as
\cite{baym1991,pines1966,bedell1993}:
\begin{equation}
\label{eq1}
\begin{split}
 \frac{\partial\delta \overrightarrow{m}_{p}}{\partial t} + \overrightarrow{\upsilon_{p}}\cdot
 \overrightarrow{\nabla}(\delta \overrightarrow{m}_{p} - \frac{\partial n_{p}^{0}}{\partial
 \varepsilon_{p}^{0}}\delta \overrightarrow{h}_{p})= \\
  -2(\overrightarrow{m}_{p} \times \delta \overrightarrow{h}_{p} + \delta \overrightarrow{m}_{p}\times
 \overrightarrow{h}_{p}^{0})+ \textit{I}[m_{p}].
\end{split}
\end{equation}
where $\delta \overrightarrow{h}_{p}= -\delta \overrightarrow{B} +
\sum_{p'}\textit{f}_{pp'}^{a} \delta \overrightarrow{m}_{p'}$ is
the fluctuation in the effective magnetic field and
$\textit{I}[m_{p}]$ is the collision integral. The equilibrium
field $h_{p}^{0}= -\textit{B} +
\sum_{p'}\textit{f}_{pp'}^{a}m_{p'}$. For the QEQ system
$\textit{B} = 0$ . We can use the relaxation time approximation
for the collision integral i.e.$\textit{I}[m_{p}] = -(\frac{\delta
m_p - \delta m }{\tau_D})$ where $\tau_D$ is the spin diffusion
lifetime where in a Fermi liquid it varies as $T^{-2}$ with the
temperature. Here we investigate the collective modes of the QEQ
system at $T\rightarrow 0K$ for which we can set
$\textit{I}[m_{p}]$ to be zero. We will discuss briefly the effect
of the finite temperature on the modes later in this paper but the
details of the calculation will be shown elsewhere. We have also
assumed that nuclear relaxation time, $T_{1}$, is sufficiently
long so that we can maintain the QEQ state for time much longer
than any diffusion time. The effect of the finite $T_{1}$ is being
studied currently.

Defining $m_{p}= -\frac{\partial n_{p}^{0}}{\partial
\varepsilon_{p}^{0}} \frac{m}{N(0)}$, and $\delta m_{p}^\pm =
\frac{-\partial n_p^0}{\partial \varepsilon_p^0}\nu^\pm_{p}$
 \cite{baym1991,pines1966}
and using $ \nu_{p}^{\pm}(r,t)=\int d^3q d\omega
\nu_{p}^{\pm}(q,\omega) e^{i(\overrightarrow{q} \cdot
\overrightarrow{r} -\omega t)} $ for the Fourier transformation of
Eq.(\ref{eq1}), we get
\begin{equation}
\label{eq5}
\begin{split}
(\omega - \overrightarrow{\upsilon_{p}}\cdot \overrightarrow{q}
\pm 2\textsl{f}_{0}^{a}m)\nu_{p}^{\pm} -
(\overrightarrow{\upsilon_{p}}\cdot \overrightarrow{q} N(0) \pm
2m)\\ \int \frac{d\Omega'}{4\pi}\textsl{f}_{pp'}^{a}\nu_{p'}^{\pm}
= -(\overrightarrow{\upsilon_{p}}\cdot \overrightarrow{q} +
\frac{2m}{N(0)}) \delta \textsl{B}^{\pm}
\end{split}
\end{equation}

\begin {figure}
%\vskip 0.15cm
\includegraphics[width = 7.50 cm,height = 3.0cm]{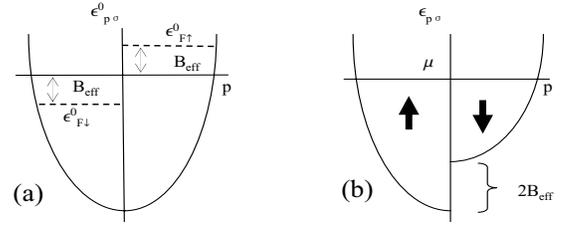}
\caption{For the QEQ and EQ systems with a majority of up spins:
(a) the change in the Fermi energy, and (b) the energy spectrum of
the up and the down spin fermions. $B_{eff} =
\frac{-mF_0^a}{N(0)}$ for the QEQ system and $B_{eff}=
\frac{m}{N(0)}= \frac{B}{1+F_0^a}$ for the EQ system.}
\end{figure}
To derive dispersion relations we set $\delta
\textsl{B}^{\pm}\rightarrow 0$.  We substitute $\nu_{p}^{\pm}$ by
its expansion in Legendre polynomials: $\nu_{p}^{\pm} = \sum_{l}
\nu_{l}^{\pm} P_{l}( \widehat{p}\cdot \widehat{q})$ and project
out the $l=0$, $l=1$, $l=2$ and $l=3$ components of
Eq.(\ref{eq5}). Because of the $\cos\theta$ term in
$\overrightarrow{\upsilon_{p}}\cdot \overrightarrow{q}$ one sees
that every moment couples to one higher moment and one lower
moment. Solving the equation for $l=3$ moment we find that
$\frac{\nu_{3}^{\pm}}{\nu_{2}^{\pm}} \lesssim \pm
\frac{q\upsilon_{F}}{2mf_{0}^{a}}$ which can be negligible for low
$q$. We also find that the exclusion of the $l=2$ moment also does
not change the important aspect of the dispersion. To see the main
features of the modes we will first work with the projection of
the $l=0$, and $l=1$ moments and truncate the distortions of the
Fermi surface, $\nu^{\pm}_{l}$, and the Landau parameters,
$F_{l}^{a}$, for $l\geq 2$.

For the $l=0$ moment, we get,
\begin{equation}
\label{eq6}
\begin{split}
\omega \nu_{0}^{\pm} - \frac{1}{3}(1 +
\frac{F_{1}^{a}}{3})q\upsilon_{F}\nu_{1}^{\pm} = 0
\end{split}
\end{equation}
which gives the net magnetization conservation law.
\begin {figure}
%\vskip 0.15cm
\includegraphics[width= 8.80 cm,height = 5.7cm]{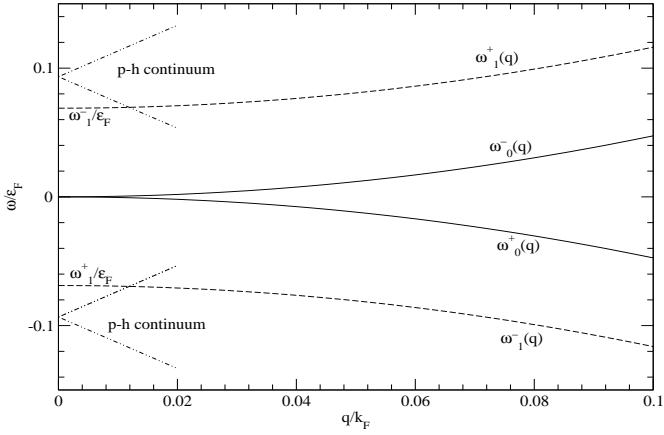}
\caption{Normalized $\omega$ for the collective modes and the p-h
continuum as a function of normalized q for the QEQ spin system.}
\end{figure}

For the $l=1$ moment, which gives the motion of the magnetization
current, we get,
\begin{equation}
\label{eq7}
\begin{split}
(\omega + \omega^{\pm}_{1})\nu_{1}^{\pm} - q\upsilon_{F}(1 +
F_{0}^{a})\nu_{0}^{\pm} = 0
\end{split}
\end{equation}
where $\omega^{\pm}_{1} = \pm \frac{2m}{N(0)}(F_{0}^{a} -
\frac{F_{1}^{a}}{3})$ is related to the internal field due to the
interacting fermions.

Solving the Eqs.(\ref{eq6},\ref{eq7}), we get,
\begin{equation}
\label{eq10}
\begin{split}
\omega(\omega + \omega^{\pm}_{1})  -
 c_{s}^{2}q^{2} = 0
\end{split}
\end{equation}
where $c_{s}^{2} = \frac{1}{3}(1 + F_{0}^{a})(1 +
\frac{F_{1}^{a}}{3})\upsilon_{F}^{2}$. We solve Eq.(\ref{eq10})
for $\omega$. The solutions are:
\begin{equation}
\label{eq11}
\begin{split}
\omega_{0}^{\pm}(q)= \frac{c_{s}^{2}q^{2}}{\omega^{\pm}_{1}}
\end{split}
\end{equation}

\begin{equation}
\label{eq12}
\begin{split}
\omega_{1}^{\pm}(q)= \omega^{\mp}_{1} +
\frac{c_{s}^{2}q^{2}}{\omega^{\mp}_{1}}
\end{split}
\end{equation}

If we take $F_{l}^{a}$ and $ \nu_{l}^{\pm} = 0$ for $l\geq 3$ and
include the projection of $l=2$, Eq.(\ref{eq10}) will be modified
to,
\begin{equation}
\label{eq13}
\begin{split}
(\omega(\omega + \omega^{\pm}_{1})- c_{s}^{2}q^2)(\omega +
\omega^{\pm}_{2}) -
 \frac{4\omega}{5} c_{s1}^{2}q^{2}  = 0
\end{split}
\end{equation}
\begin {figure}
%\vskip 0.15cm
\includegraphics[width= 8.5 cm,height = 5.7cm]{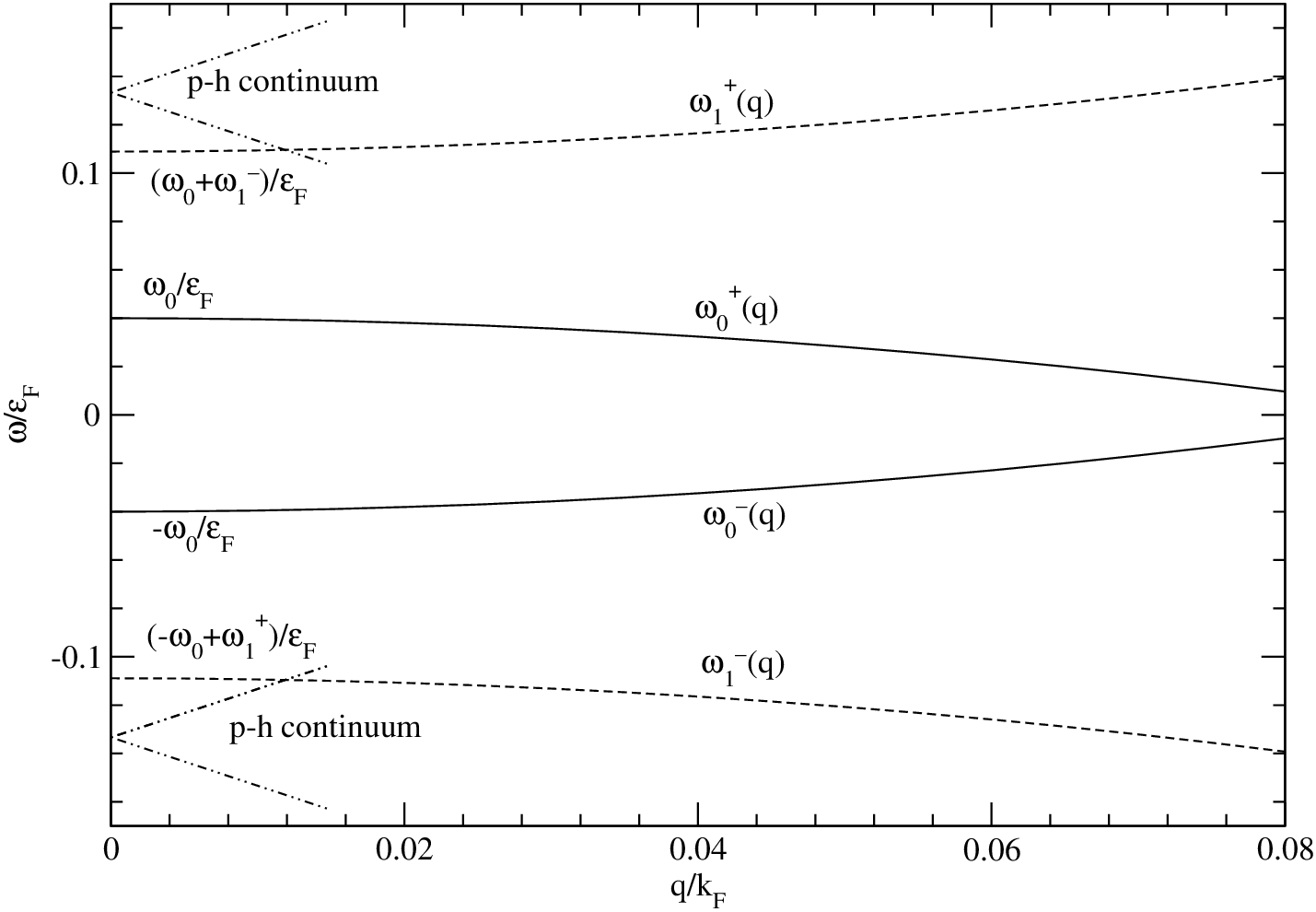}
\caption{Normalized $\omega$ for the collective modes and the p-h
continuum as a function of normalized q in the EQ spin system.}
\end{figure}
where $c_{s1}^{2} = \frac{1}{3}(1 + \frac{F_{1}^{a}}{3})(1 +
\frac{F_{2}^{a}}{3})\upsilon_{F}^{2}$. From Eq.(\ref{eq13}) we can
show that the contribution of order $q^{2}$ to the dispersion
relations will be only to $\omega_{1}^{\pm}(q)$, which now reads
as,
\begin{equation}
\label{eq14}
\begin{split}
 \omega^{\pm}_{1}(q)= \omega^{\mp}_{1} +
 \frac{c_{s}^{2}q^{2}}{\omega^{\mp}_{1}}\mp \frac{4N(0)\upsilon_{F}^{2}}{30m}( \frac{3}{F_{1}^{a}} + 1)q^{2}
\end{split}
\end{equation}
For the EQ system, Eq.(\ref{eq5}) will have the form
\begin{equation}
\label{eq15}
\begin{split}
(\omega \mp \omega_{0} - \overrightarrow{\upsilon_{p}}\cdot
\overrightarrow{q} \pm 2\textsl{f}_{0}^{a}m)\nu_{p}^{\pm} -
(\overrightarrow{\upsilon_{p}}\cdot \overrightarrow{q} N(0) \pm
2m)\\ \int \frac{d\Omega'}{4\pi}\textsl{f}_{pp'}^{a}\nu_{p'}^{\pm}
= -(\overrightarrow{\upsilon_{p}}\cdot q + \frac{2m}{N(0)}) \delta
\textsl{B}^\pm
\end{split}
\end{equation}
which modifies the Eqs.(\ref{eq11}) and (\ref{eq12}) as follows:
\begin{equation}
\label{eq16}
\begin{split}
\omega_{0}^{\pm}(q)= \pm \omega_{0} +
 \frac{c_{s}^{2}q^{2}}{\omega^{\pm}_{1}}
\end{split}
\end{equation}
\label{eq17}
\begin{equation}
\begin{split}
\omega_{1}^{\pm}(q)= \pm \omega_{0} + \omega^{\mp}_{1} +
\frac{c_{s}^{2}q^{2}}{\omega^{\mp}_{1}}
\end{split}
\end{equation}
where $m=\frac{N(0)}{1+F_{0}^{a}}B$. We mention here, for the
purpose of later use,
 that the particle-hole(p-h) continuum for the QEQ
and the EQ systems  are given by $\omega_{ph}^{\pm}= \mp 2mf_0^a +
\overrightarrow{q}.\overrightarrow{v}$ and $\omega_{ph}^{\pm}= \pm
\omega_0 \mp 2mf_0^a + \overrightarrow{q}.\overrightarrow{v}$,
respectively.

 We evaluated the dispersion relations of the collective modes and the p-h continuum for $10\%$ polarized
 QEQ and EQ spin
 systems using the Fermi liquid parameters suitable for a polarized
 liquid
 helium at zero pressure\cite{greywall1982}. For the EQ system
 $\frac{\omega_{0}}{\varepsilon_{F}}=0.04$ corresponds to $10\%$ polarization.
 The results have been presented in Figs.(2,3).

In Fig.2, we see that the QEQ system has one gapless and one
gapped collective mode. In principle, we will have $l$ gapped
modes if we include up to $l^{th}$ term of the expansion of
$f_{pp'}^{s,a}$ and $\nu_p^\pm$ in the Landau kinetic equation. We
have shown one of those gapped modes in Fig. 2. The gapless mode
has a dispersion relation similar to that of the Goldstone mode of
a ferromagnetic system\cite{blagoev2001}. The gapless mode is
related to the precessional mode of the magnetization in the
absence of an external field. This fact can be understood by
setting $ q=0$ in Eq.(\ref{eq6}), which gives $\omega \nu_0^{\pm}=
0$, and for $\nu_0^{\pm} \neq 0$ we get $\omega = 0$ and hence a
gapless mode. The gapped mode originates from the oscillation of
the spin current,$\overrightarrow{j_\sigma} \propto \nu_1$. From
Eq.(\ref{eq7}) with $q=0$ we have $(\omega +
\omega_1^\pm)\nu_1^\pm = 0$, and for $\nu_1^\pm \neq 0$ we find
$\omega = \omega_1^\mp$. Thus the gapped mode is related to the
precession of the spin current around the internal field due to
the interacting fermions. We also see that these collective modes
are well separated from the p-h continuum at low q and hence will
be propagating modes. If we did not include $f_1^a$ in the kinetic
equation, the current mode would lie inside the p-h continuum, and
hence would be Landau damped. This shows the importance of
including $f_1^a$ in the kinetic equation. In Fig.3, we see that
both the collective modes of the EQ system are also propagating.
But contrary to the QEQ system, none of the modes are gapless. The
uniform mode in the EQ system corresponds to the precession of the
magnetization around the external field at the Larmor frequency,
$\pm\omega_0$. The current mode corresponds to the precession of
the spin current and is gapped with $\pm\omega_0 + \omega_1^\mp$.

The presence of the $f_{1}^{a}$ in the kinetic equation has an
interesting consequence in the dynamical structure function,
$S(q,\omega)$. From Eq.(\ref{eq5}), we find an expression for the
dynamical spin susceptibility using $\chi^\pm(q,\omega) =
\frac{-N(0)\nu_0^\pm}{\delta B^\pm}$\cite{baym1991, pines1966} and
used it to obtain
\begin{equation}
 S^\pm(q,\omega)= \frac{-1}{2\pi}(Im[\chi^\pm(q,\omega)]-
Im[\chi^\pm(q,-\omega)])
\end{equation}
to study the frequency dependence of $S^\pm(q,\omega)$ and the
f-sum rule. For experimental purpose $S^\pm(q,\omega)$ is defined
in relation to the scattering cross section as
$\frac{d^2\sigma}{d\Omega d\omega}= \sigma_0
S^\pm(q,\omega)$\cite{nozieres1997} where $\sigma_0$ contains all
the elements related to the probe particle interaction and
$S^\pm(q,\omega)$ characterizes the many-body system. The f-sum
rule states that $\int \omega S^\pm(q,\omega) d\omega =
(1+\frac{F_{1}^{a}}{3}) \frac{n q^{2}}{2m^{*}}$
\cite{leggett1966}, where $n$ is the total density, and $m^{*}$ is
the effective mass. For the QEQ system we see that the gapless and
the gapped collective modes exhaust the sum rule if we include the
$f_{1}^{a}$ in the kinetic equation. In this case the p-h
contribution to the sum will be proportional to $q^{4}$. But if we
did not include $f_{1}^{a}$ the p-h continuum would also have a
contribution proportional to $q^{2}$. The gapped mode, which comes
out of the p-h continuum due to the inclusion of $f_{1}^{a}$ in
the kinetic equation, takes out most of the spectral weight of the
p-h continuum contribution to the sum rule. We observe a similar
behavior for the EQ phase, as was found earlier in the
ferromagnetic system\cite{blagoev2001}.

In addition to the dynamical consequences of the gapless mode
there may be interesting thermodynamic properties that can be
observed in the QEQ system. Perhaps the most dramatic would be a
$T^{3/2}$ contribution to the specific heat. Whereas, for the EQ
system, because all of the transverse collective modes are gapped,
the specific heat contribution would decreases exponentially with
the decrease in temperature.

As mentioned earlier, to study the effect of the temperature on
the collective modes, we solved the Landau kinetic equation using
the relaxation-time approximation for the collision integral. We
find that the real part of the dispersion relations remains
dominant at low temperature. The dispersion relation shows a
transition of the spin system from the collision less region to a
hydrodynamic region as has been discussed in the
literature\cite{baym1991}.

In summary using the Landau Fermi liquid theory we explored the
thermodynamics and the dynamics of a paramagnetic spin system
which is out of equilibrium. We report for the first time the
prediction of a gapless collective mode in such systems. We
discuss the difference in the nature of the collective modes in
the QEQ and the EQ systems. We point out the importance of
including the $f_{1}^{a}$ Landau parameter in the kinetic
equation. We believe that some of the qualitative features of the
collective modes, in particular the gapless mode, will also be
seen in systems far from equilibrium. We want to emphasize that
this study is applicable in a wide variety of interesting
materials.

This work is done with the support of grant DOE/DEFG0297ER45636.
One of the authors KSB, would like to thank the Aspen Center for
Physics and the KITP in Santa Barbara for their hospitality. We
would also like to thank Dr. Krastan Blagoev and Raul Chura for
valuable discussion and advice.

\bibliographystyle{apsrev}
\bibliography{references}

\begin{thebibliography}{21}
\expandafter\ifx\csname natexlab\endcsname\relax\def\natexlab#1{#1}\fi
\expandafter\ifx\csname bibnamefont\endcsname\relax
  \def\bibnamefont#1{#1}\fi
\expandafter\ifx\csname bibfnamefont\endcsname\relax
  \def\bibfnamefont#1{#1}\fi
\expandafter\ifx\csname citenamefont\endcsname\relax
  \def\citenamefont#1{#1}\fi
\expandafter\ifx\csname url\endcsname\relax
  \def\url#1{\texttt{#1}}\fi
\expandafter\ifx\csname urlprefix\endcsname\relax\def\urlprefix{URL }\fi
\providecommand{\bibinfo}[2]{#2}
\providecommand{\eprint}[2][]{\url{#2}}

\bibitem[{\citenamefont{Bouchiat et~al.}(1960)\citenamefont{Bouchiat, Carver,
  and Varnum}}]{bouchiat1960}
\bibinfo{author}{\bibfnamefont{M.}~\bibnamefont{Bouchiat}},
  \bibinfo{author}{\bibfnamefont{T.}~\bibnamefont{Carver}}, \bibnamefont{and}
  \bibinfo{author}{\bibfnamefont{C.}~\bibnamefont{Varnum}},
  \bibinfo{journal}{Phys. Rev. Lett.} \textbf{\bibinfo{volume}{5}},
  \bibinfo{pages}{373} (\bibinfo{year}{1960}).

\bibitem[{\citenamefont{Colegrove et~al.}(1963)\citenamefont{Colegrove,
  Schearer, and Walters}}]{colegrove1963}
\bibinfo{author}{\bibfnamefont{F.}~\bibnamefont{Colegrove}},
  \bibinfo{author}{\bibfnamefont{L.}~\bibnamefont{Schearer}}, \bibnamefont{and}
  \bibinfo{author}{\bibfnamefont{G.}~\bibnamefont{Walters}},
  \bibinfo{journal}{Phys. Rev.} \textbf{\bibinfo{volume}{132}},
  \bibinfo{pages}{2561} (\bibinfo{year}{1963}).

\bibitem[{\citenamefont{Leggett}(2001)}]{leggett2001}
\bibinfo{author}{\bibfnamefont{A.~J.} \bibnamefont{Leggett}},
  \bibinfo{journal}{Rev. Mod. Phys.} \textbf{\bibinfo{volume}{73}},
  \bibinfo{pages}{307} (\bibinfo{year}{2001}).

\bibitem[{\citenamefont{Ketterle}(2002)}]{ketterle2002}
\bibinfo{author}{\bibfnamefont{W.}~\bibnamefont{Ketterle}},
  \bibinfo{journal}{Rev. Mod. Phys.} \textbf{\bibinfo{volume}{74}},
  \bibinfo{pages}{1131} (\bibinfo{year}{2002}).

\bibitem[{\citenamefont{Zwierlein et~al.}(2005)\citenamefont{Zwierlein,
  Schirotzek, Schunck, and Ketterle}}]{martin2005}
\bibinfo{author}{\bibfnamefont{M.~W.} \bibnamefont{Zwierlein}},
  \bibinfo{author}{\bibfnamefont{A.}~\bibnamefont{Schirotzek}},
  \bibinfo{author}{\bibfnamefont{C.~H.} \bibnamefont{Schunck}},
  \bibnamefont{and} \bibinfo{author}{\bibfnamefont{W.}~\bibnamefont{Ketterle}},
  \bibinfo{journal}{cond-mat/0511197 and the references therein}
  (\bibinfo{year}{2005}).

\bibitem[{\citenamefont{Zutic et~al.}(2004)\citenamefont{Zutic, Fabian, and
  Sarma}}]{zutic2004}
\bibinfo{author}{\bibfnamefont{I.}~\bibnamefont{Zutic}},
  \bibinfo{author}{\bibfnamefont{J.}~\bibnamefont{Fabian}}, \bibnamefont{and}
  \bibinfo{author}{\bibfnamefont{S.~D.} \bibnamefont{Sarma}},
  \bibinfo{journal}{Rev. Mod. Phys.} \textbf{\bibinfo{volume}{76}},
  \bibinfo{pages}{323, and the references therein} (\bibinfo{year}{2004}).

\bibitem[{\citenamefont{Cho}(2005)}]{cho2005}
\bibinfo{author}{\bibfnamefont{A.}~\bibnamefont{Cho}},
  \bibinfo{journal}{Science} \textbf{\bibinfo{volume}{310}},
  \bibinfo{pages}{1892} (\bibinfo{year}{2005}).

\bibitem[{\citenamefont{Fernandez-Rossier
  et~al.}(2004)\citenamefont{Fernandez-Rossier, Piermarocchi, Chen, MacDonald,
  and Sham}}]{fernandez2004}
\bibinfo{author}{\bibfnamefont{J.}~\bibnamefont{Fernandez-Rossier}},
  \bibinfo{author}{\bibfnamefont{C.}~\bibnamefont{Piermarocchi}},
  \bibinfo{author}{\bibfnamefont{P.}~\bibnamefont{Chen}},
  \bibinfo{author}{\bibfnamefont{A.}~\bibnamefont{MacDonald}},
  \bibnamefont{and} \bibinfo{author}{\bibfnamefont{L.}~\bibnamefont{Sham}},
  \bibinfo{journal}{Phys. Rev. Lett.} \textbf{\bibinfo{volume}{93}},
  \bibinfo{pages}{127201} (\bibinfo{year}{2004}).

\bibitem[{\citenamefont{Mal'shukov et~al.}(2005)\citenamefont{Mal'shukov, Tang,
  Chu, and Chao}}]{malshukov2005}
\bibinfo{author}{\bibfnamefont{A.}~\bibnamefont{Mal'shukov}},
  \bibinfo{author}{\bibfnamefont{C.}~\bibnamefont{Tang}},
  \bibinfo{author}{\bibfnamefont{C.}~\bibnamefont{Chu}}, \bibnamefont{and}
  \bibinfo{author}{\bibfnamefont{K.}~\bibnamefont{Chao}},
  \bibinfo{journal}{Phys. Rev. Lett.} \textbf{\bibinfo{volume}{95}},
  \bibinfo{pages}{107203} (\bibinfo{year}{2005}).

\bibitem[{\citenamefont{Tserkovnyak et~al.}(2000)\citenamefont{Tserkovnyak,
  Brataas, and Bauer}}]{yaroslav2002}
\bibinfo{author}{\bibfnamefont{Y.}~\bibnamefont{Tserkovnyak}},
  \bibinfo{author}{\bibfnamefont{A.}~\bibnamefont{Brataas}}, \bibnamefont{and}
  \bibinfo{author}{\bibfnamefont{G.~E.~W.} \bibnamefont{Bauer}},
  \bibinfo{journal}{Phys. Rev. B} \textbf{\bibinfo{volume}{66}},
  \bibinfo{pages}{224403} (\bibinfo{year}{2000}).

\bibitem[{\citenamefont{Candela et~al.}(1994)\citenamefont{Candela, Hayden, and
  Nacher}}]{candela1994}
\bibinfo{author}{\bibfnamefont{D.}~\bibnamefont{Candela}},
  \bibinfo{author}{\bibfnamefont{M.}~\bibnamefont{Hayden}}, \bibnamefont{and}
  \bibinfo{author}{\bibfnamefont{P.~J.} \bibnamefont{Nacher}},
  \bibinfo{journal}{Phys. Rev. Lett.} \textbf{\bibinfo{volume}{73}},
  \bibinfo{pages}{2587, and the references therein} (\bibinfo{year}{1994}).

\bibitem[{\citenamefont{Castaing and Nozieres}(1979)}]{castaing1979}
\bibinfo{author}{\bibfnamefont{B.}~\bibnamefont{Castaing}} \bibnamefont{and}
  \bibinfo{author}{\bibfnamefont{P.}~\bibnamefont{Nozieres}},
  \bibinfo{journal}{J. Phys.(Paris)} \textbf{\bibinfo{volume}{40}},
  \bibinfo{pages}{257} (\bibinfo{year}{1979}).

\bibitem[{\citenamefont{Chapellier et~al.}(1979)\citenamefont{Chapellier,
  Fossati, and Rasmussen}}]{chapellier1979}
\bibinfo{author}{\bibfnamefont{M.}~\bibnamefont{Chapellier}},
  \bibinfo{author}{\bibfnamefont{G.}~\bibnamefont{Fossati}}, \bibnamefont{and}
  \bibinfo{author}{\bibfnamefont{F.}~\bibnamefont{Rasmussen}},
  \bibinfo{journal}{Phys. Rev. Lett.} \textbf{\bibinfo{volume}{42}},
  \bibinfo{pages}{904} (\bibinfo{year}{1979}).

\bibitem[{\citenamefont{Bedell}(1994)}]{bedell1993}
\bibinfo{author}{\bibfnamefont{K.}~\bibnamefont{Bedell}}, in
  \emph{\bibinfo{booktitle}{Modern Perspectives in Many-Body Physics:
  Proceedings:Proceedings of the Sixth Physics Summer School , 1993}}, edited
  by \bibinfo{editor}{\bibfnamefont{M.~P.} \bibnamefont{Das}} \bibnamefont{and}
  \bibinfo{editor}{\bibfnamefont{J.}~\bibnamefont{Mahanty}},
  \bibinfo{organization}{The Australian National University}
  (\bibinfo{publisher}{World Scientific}, \bibinfo{year}{1994}), pp.
  \bibinfo{pages}{119--154}.

\bibitem[{\citenamefont{Sanchez-Castro
  et~al.}(1989)\citenamefont{Sanchez-Castro, Bedell, and Wiegers}}]{carlos1989}
\bibinfo{author}{\bibfnamefont{C.~R.} \bibnamefont{Sanchez-Castro}},
  \bibinfo{author}{\bibfnamefont{K.~S.} \bibnamefont{Bedell}},
  \bibnamefont{and} \bibinfo{author}{\bibfnamefont{S.~A.~J.}
  \bibnamefont{Wiegers}}, \bibinfo{journal}{Phys. Rev. B}
  \textbf{\bibinfo{volume}{40}}, \bibinfo{pages}{437} (\bibinfo{year}{1989}).

\bibitem[{\citenamefont{Baym and Pathick}(1991)}]{baym1991}
\bibinfo{author}{\bibfnamefont{G.}~\bibnamefont{Baym}} \bibnamefont{and}
  \bibinfo{author}{\bibfnamefont{C.}~\bibnamefont{Pathick}},
  \emph{\bibinfo{title}{Landau Fermi-liquid Theory}} (\bibinfo{publisher}{New
  York, Wiley}, \bibinfo{year}{1991}).

\bibitem[{\citenamefont{Pines and Nozieres}(1966)}]{pines1966}
\bibinfo{author}{\bibfnamefont{D.}~\bibnamefont{Pines}} \bibnamefont{and}
  \bibinfo{author}{\bibfnamefont{P.}~\bibnamefont{Nozieres}},
  \emph{\bibinfo{title}{The theory of quantum liquids, vol.I}}
  (\bibinfo{publisher}{Advanced Book Classics}, \bibinfo{year}{1966}).

\bibitem[{\citenamefont{Greywall}(1982)}]{greywall1982}
\bibinfo{author}{\bibfnamefont{D.~S.} \bibnamefont{Greywall}},
  \bibinfo{journal}{Phys. Rev. B} \textbf{\bibinfo{volume}{27}},
  \bibinfo{pages}{2747} (\bibinfo{year}{1982}).

\bibitem[{\citenamefont{Blagoev and Bedell}(2001)}]{blagoev2001}
\bibinfo{author}{\bibfnamefont{K.~B.} \bibnamefont{Blagoev}} \bibnamefont{and}
  \bibinfo{author}{\bibfnamefont{K.~S.} \bibnamefont{Bedell}},
  \bibinfo{journal}{Phil. Mag. Lett.} \textbf{\bibinfo{volume}{81}},
  \bibinfo{pages}{511} (\bibinfo{year}{2001}).

\bibitem[{\citenamefont{Nozieres}(1997)}]{nozieres1997}
\bibinfo{author}{\bibfnamefont{P.}~\bibnamefont{Nozieres}},
  \emph{\bibinfo{title}{Theory of Intracting Fermi System}}
  (\bibinfo{publisher}{Advanced Book Classics}, \bibinfo{year}{1997}).

\bibitem[{\citenamefont{Leggett}(1966)}]{leggett1966}
\bibinfo{author}{\bibfnamefont{A.~J.} \bibnamefont{Leggett}},
  \bibinfo{journal}{Phys. Rev.} \textbf{\bibinfo{volume}{147}},
  \bibinfo{pages}{119} (\bibinfo{year}{1966}).

\end{thebibliography}
\end{document}